\documentclass[11pt]{article}
\newcommand{\ket}[1]{\left | #1 \right \rangle}
\newcommand{\bra}[1]{\left \langle #1 \right |}

\def\openone{\leavevmode\hbox{\small1\kern-3.8pt\normalsize1}}
\def\RR{{\rm I\kern-.2emR}}
\def\tr{{\rm tr}\; }
\def\ce{{\cal E}}
\def\cd{{\cal D}}

\def\on{^{\otimes N}}
\def\pn{^{(N)}}

\def\rhoprime{\rho^{\,\prime}}

\newcommand{\proj}[1]{\ket{#1}\!\bra{#1}}

\newcommand{\inner}[2]{ \langle #1 | #2 \rangle}

\newcommand{\beq}{\begin{equation}}
\newcommand{\eeq}{\end{equation}}
\newcommand{\beqa}{\begin{eqnarray}}
\newcommand{\eeqa}{\end{eqnarray}}
\newcommand{\QED}{\hspace*{\fill}\mbox{\rule[0pt]{1.5ex}{1.5ex}}}

\begin{document}
\begin{center}
{\Large\bf On quantum coding for ensembles of mixed states}\\
\bigskip
{\normalsize Howard Barnum$^\dagger$, Carlton M. Caves$^\S$,
Christopher A. Fuchs$^\|$,\\
Richard Jozsa$^\dagger$, and Benjamin Schumacher$^+$}\\
\bigskip
{\small\it $^\dagger$Department of Computer Science, 
University of Bristol,\\ Merchant Venturers Building, Bristol BS8 1UB U.K. \\
$^\S$Department of Physics and Astronomy, University of New
Mexico,\\ Albuquerque, NM 87131 U.S.A. \\
$^\|$Los Alamos National Laboratory, MS-B285, Los Alamos, NM 
87545 U.S.A.\\
$^+$Department of Physics, Kenyon College, Gambier, OH 43022
U.S.A.}
\\[4mm]
\date{today}
\end{center}

\begin{abstract}
We consider the problem of optimal asymptotically faithful
compression for ensembles of mixed quantum states. Although the
optimal rate is unknown, we prove upper and lower bounds and
describe a series of illustrative examples of compression of mixed
states. We also discuss a classical analogue of the problem.
\end{abstract}

\section{Introduction}

The emergence of potentially useful theoretical protocols for using 
quantum states in cryptography and quantum computation has increased the 
theoretical (and perhaps ultimately practical) importance of questions 
about how quantum states can be compressed, transmitted across noisy or 
low-dimensional channels, and recovered, and otherwise manipulated in a
fashion analogous to classical information.  Most of the work done on these 
matters, beginning with \cite{Schumacher95a}, 
has focused on the manipulation 
of pure states, with mixed states appearing only in intermediate stages, as 
the result of noise.  An exception is \cite{Barnum96a}, which considered
the copying or broadcasting of mixed states.
When mixed states have appeared as states to be 
transmitted, it has usually been required that their potential entanglement
with some reference system be preserved,
as in \cite{Barnum98a}.  This focuses attention again on 
a {\em pure\/} state, the entangled state of system and reference system.
As discussed further in 
\cite{Barnum2000a} there is a close relation between entanglement 
transmission and the transmission of pure states of the system itself.

In the present paper, we consider the compression or transmission of mixed 
states, {\em without\/} any requirement that their entanglement or correlation
with other systems be preserved.  There might seem to be good reason to 
confine oneself to pure-state transmission, since mixed states, considered 
apart from any potential entanglement with other systems, might not seem 
particularly useful.  This may be why the classical analogue of the problem we 
consider in this paper---the transmission of probability distributions---has 
not, to our knowledge, been previously studied.  Game theory is perhaps the 
first situation that springs to mind in which one might wish to produce a 
mixed state intentionally, given that all pure states of which it may be 
viewed as a mixture are available, since it is well known to game theorists 
that mixed strategies may be better than any of their component pure 
strategies in important situations \cite{vonNeumann44a,Luce57a}.  
Thus a ``practical'' application of 
mixed-state compression might be the compression of mixed strategies, where 
the ``decoding'' is done by the player playing the strategy or someone who 
shares his goals.  In cryptographic applications (closely related to game 
theory, of course) and also in probabilistic classical algorithms, there may 
be a use for randomness and an interest in compressing it for efficient
storage or transmission.  Indeed, quantum computation can enable 
more-efficient-than-classical sampling from probability distributions 
\cite{Grover99a,Bernstein97a}; there may be relations between these ideas 
and the work reported here.

The problem of optimal compression for ensembles of pure quantum
states has been solved \cite{Schumacher95a,Jozsa94d,Barnum96b}, but
for sources of mixed states the minimal resources are unknown.
This question has also been considered by M. Horodecki in
\cite{Horodecki98a,Horodecki99a}.  In this paper, we consider several
variants of the question, depending on the fidelity criteria and 
encoding/decoding procedures used.  Sections 2 and 3 present the
problem, in variants depending on whether or not the encoder/compressor knows
the identity of each state and can use it to help encode, and depending
on whether, in a block-coding setting, a marginal (``local'') 
or total (``global'') fidelity 
criterion is used;  Section 4 considers relations between these variants of
the problem, in general and for the special case of pure states.  
Section 5 discusses the fact that the entropy of a source ensemble's 
average density operator provides (as in the pure-state case) 
an upper bound on the rate at which qubits must be used to represent
the source.  We also show that under the global fidelity criterion,
if decodings are required to be unitary, this is actually 
the optimal rate.
Section 6
formulates a classical version of the problem, which we have not seen 
treated in classical information theory, and discusses examples.  In
Section 7, we show with several examples
that in contrast to the pure state case, it is possible with {\em general} 
decodings to 
compress to below the entropy of the average density operator. 
This section also introduces a useful 
preparation-visible 
technique, that of compression by purifications, which we show does better 
than our classical methods for some of the classical mixed-state compression
problems considered in Section 6.  Finally, in Section 8
we show that 
the Holevo quantity $S(\sum_i p_i \sigma_i) - \sum_i p_i S(\sigma_i)$ for
an ensemble gives a lower bound on the qubit rate required to represent a 
source.  (A different proof is given in \cite{Horodecki98a}.)
We do not know whether this lower bound is attainable in general.

\section{Formulation of the Problem}

In this paper, $S(\rho)$ will always denote the von Neumann entropy
of a density matrix $\rho$ and $H(p_1,...,p_n)$ will denote
the Shannon entropy of a probability distribution $p_1,...,p_n$.
In both cases logarithms are taken to base $2$:
\beqa
S(\rho) := - \tr (\rho  {~\rm log}_2~\rho) \nonumber \\
H(p_1,...,p_n) := - \sum_i p_i {~\rm log}_2~ p_i \nonumber 
\eeqa

Let $\rho_1,\ldots,\rho_n $ be a list of (possibly mixed)
$d$-dimensional quantum states. Each state is assigned a prior
probability $p_1, \ldots, p_n$ respectively. We refer to such a
list as a source or ensemble of signal states, denoted by 
$E= \{ p_i,\rho_i \}$. Alice is fed an unending sequence of these signal
states, with each successive state chosen randomly and independently 
from $E$. At time $N$ she will have the total state 
$\sigma_N = \rho_{i_1} \otimes \cdots \otimes \rho_{i_N}$ with
probability $p_{i_1} p_{i_2} \cdots p_{i_N}$.

Alice wants to perform either of the following tasks (which are
equivalent for our considerations):
\begin{itemize}
\item{Communication: Alice wants to send the signals to Bob using a minimum
number of qubits/signal so that Bob can reconstruct long sequences
with ``arbitrarily high fidelity''. This involves a ``coding procedure''
for Alice and a ``decoding procedure'' for Bob (cf.~later discussion
for the precise meaning of all these terms).}
\item{Storage: Alternatively, Alice wants to
store the signals as efficiently as possible. In this interpretation
the coding procedure is used for putting signals into storage, and
the decoding procedure for reconstituting them.}
\end{itemize}
\noindent We distinguish two fundamental situations for Alice:

\begin{itemize}
\item{Preparation-blind (blind): Alice is not given the identity
of the individual (generally nonorthogonal) signal states
(she knows only their prior distribution).}

\item{Preparation-visible (visible): Alice is given the identity of the
individual signal states (as well as their prior distribution).
Indeed, in this case we may assume that she is simply provided with a
sequence of the {\em names\/} of the states and she may prepare the states
herself if she wishes.}
\end{itemize}

Note that in the blind case, Alice is being fed essentially quantum
information, whereas in the visible case she is getting entirely classical
information. In both cases, however, Bob on decoding is not required to 
identify the actual signal states, but only to produce high fidelity 
representatives of the correct sequence of states. Hence, even in the 
visible case, the problem is not one of classical coding/information 
theory.  The visible case (for pure states) occurs, for example, in 
quantum cryptographic protocols (e.g. BB84 \cite{Bennett84a} and 
B92 \cite{Bennett92a}), where the sender (Alice) is also the state preparer.

\section{Coding/Decoding Schemes and their Fidelity}

Let ${\cal H}_d$ denote the space of all $d$-dimensional states.
Given any physical system in state $\rho$, quantum mechanics
allows only the following three types of operations: 

\begin{itemize}
\begin{item}
(OP1) A unitary transformation, $\rho \rightarrow U\rho U^{\dagger}$ ($U$
unitary).
\end{item}
\begin{item}
(OP2) Inclusion of an ancilla in a standard state $\rho_0$ (independent of 
$\rho$), $\rho \rightarrow \rho \otimes \rho_0$. 
\end{item}
\begin{item}
(OP3) Discarding a subsystem (when $\rho$ is a state of a composite system 
$AB$), $\rho_{AB} \rightarrow {\rm tr}_B (\rho_{AB})$.  
\end{item}
\end{itemize}
\noindent
Note that (OP2) and (OP3) change the value of $d$.

Consider any length-$N$ string of input states (given either
visible or blind):
\begin{equation}
\sigma_N = \rho_{i_1} \otimes \cdots \otimes \rho_{i_N} \in 
{\cal H}_d^{\otimes N}\;,\label{inp}
\end{equation}
\begin{equation}
{\rm Prob}(\sigma_N) = p_{i_1} \cdots p_{i_N}\;. \label{prob}
\end{equation}
A coding/decoding scheme, using $q$ qubits/signal, is defined by
the following requirements, which are to be specified for all
sufficiently large $N$. 

\begin{itemize}
\item{Blind coding: Alice's coding procedure,
if blind, is any specified sequence of the above three operations
applied to $\sigma_N$, giving a final state $\omega_N$ within the
required resources of $q$ qubits/signal (i.e., in $2^{qN}$ dimensions).
Mathematically, any such sequence of operations corresponds to a
completely positive, trace-preserving map $\cal C$ on the density
operator, i.e., $\omega_N = {\cal C} 
( \rho_{i_1} \otimes \cdots \otimes \rho_{i_N} )$,
and any completely positive, trace-preserving map corresponds to such
a sequence of operations.}
\item{Visible coding: Alice's coding procedure, if visible,
corresponds to an {\em arbitrary\/} assignment of a state $\omega_N
\in {\cal H}_{2^{qN}} $ to each $\sigma_N$; i.e., Alice can build any
state she pleases as the coded version of the input string.}
\item{Finally, Bob's decoding (analogous to blind coding) is any
sequence of the above three operations applied to the coded state,
yielding a state $\tilde{\sigma}_N$ of $N$ $d$-dimensional systems.
Thus decoding is a completely positive, trace-preserving map from 
${\cal H}_{2^{qN}}$ to ${\cal H}_{d^N}$.}
\end{itemize}

Let us write Bob's decoded state, produced by coding followed by decoding 
of $\sigma_N = \rho_{i_1} \otimes \cdots \otimes \rho_{i_N}$, as
$\tilde{\sigma}_N = \tilde{\sigma}_{i_1 \ldots i_N}$.
Let
\begin{equation}
\tilde{\rho}_k = 
\left\{
\begin{array}{l} 
\mbox{trace of $\tilde{\sigma}_{i_1,...,i_N}$ over all} \\
\mbox{signal spaces
except the $k$-th}
\end{array}
\right\}
~k = 1,...N\;
\label{tr}
\end{equation}
be the reduced state in the $k^{\rm th}$ signal position after
coding and decoding; i.e., $\tilde{\rho}_k$ is the decoded version
of the $k^{th}$ transmitted state $\rho_{i_k}$. Let
\begin{equation}
F(\rho_1 ,\rho_2 ) = \left( \mbox{trace}(\rho_1^{1/2}\rho_2
\rho_1^{1/2})^{1/2}
  \right)^2  \label{uhl}
\end{equation}
denote the Bures-Uhlmann fidelity function \cite{Bures69a,
Uhlmann76a, Jozsa94b}. The coding/decoding scheme has fidelity
$1-\epsilon$ if it satisfies the following fidelity requirement:
There is an $N_0$ such that for all $N>N_0$,
\begin{equation}
\sum_{\sigma_N} {\rm Prob}(\sigma_N ) \prod_{k=1}^{N}
F(\rho_{i_k}, \tilde{\rho}_k ) > 1-\epsilon \hspace{2cm}
\mbox{(LOCAL-FID),}
\end{equation}
Note that high fidelity according to (LOCAL-FID) allows entanglement 
to be introduced between output signal states, even though there was 
no entanglement in the input (\ref{inp}), which was taken to be a
product state. This is because we examine $\tilde{\sigma}_N$ only
through its partial traces (\ref{tr}), thus reducing the state to 
each position separately. In view of this we might consider a stronger
fidelity criterion (GLOBAL-FID), which replaces (LOCAL-FID) by
\begin{equation}
\sum_{\sigma_N} {\rm Prob}(\sigma_N )\, F(\sigma_N
,\tilde{\sigma}_N )
> 1-\epsilon \hspace{2cm}  \mbox{(GLOBAL-FID).}
\end{equation}
For $\epsilon$ tending to zero, this eliminates extraneous
entanglements in the output sequence.  Note that in a
continuously varying situation with $\epsilon$ tending to zero,
(GLOBAL-FID) implies (LOCAL-FID) because (GLOBAL-FID) will require
that $\tilde{\sigma}_N$ become arbitrarily close to $\sigma_N$
and, hence, $F(\rho_{i_k} , \tilde{\rho}_k )\rightarrow 1$ for each
$k$, too.
\\[2mm] 
{\bf Example 1}. Alice wants to send Bob $\sigma_2
= \frac{1}{2}I \otimes \frac{1}{2}I$, so we may take the decoded
state to be $\tilde{\sigma_2 }= \frac{1}{2}I \otimes \frac{1}{2}I$
satisfying (LOCAL-FID) and (GLOBAL-FID) with $\epsilon = 0$ or
\[ \tilde{\sigma_2} = 
\frac{1}{2}\Bigl(\ket{0}\otimes\ket{0} + \ket{1}\otimes\ket{1}\Bigr) 
\Bigl(\bra{0}\otimes\bra{0} + \bra{1}\otimes\bra{1}\Bigr) \;, \]
satisfying (LOCAL-FID) with $\epsilon =0$ but not (GLOBAL-FID).
$\QED$

We will generally adopt the fidelity requirement (GLOBAL-FID) 
in the following.  If it is important that the signal states 
remain uncorrelated, (GLOBAL-FID) is the appropriate criterion; 
otherwise it may be too strong.
\\[2mm]
{\bf Remark 2}. (LOCAL-FID) has the following awkward feature:
If we have a very high fidelity coding/decoding scheme according
to (LOCAL-FID) and we repeatedly apply it to a long string, 
$\sigma_N \rightarrow \tilde{\sigma}_N \rightarrow \tilde{\tilde{\sigma}}_N$, 
then we will not necessarily preserve high fidelity in the sequence of
reduced states. This is because though $\sigma_N$ and $\tilde{\sigma}_N$
have essentially the same reduced states at each position, globally
they can be very different states (cf.~Example 1). Since the coding 
scheme is generally a {\em block\/}-coding scheme, it uses the global 
input state and will work well only if this global state is a 
{\em product\/} state as in (\ref{inp}). Hence $\tilde{\tilde{\sigma}}_N$ 
will not generally have the correct reduced states. 
$\QED$

From the above precise formulations of the notions of coding,
decoding, and fidelity, we obtain a well defined mathematical
problem.
\\[2mm] 
{\bf Problem~3.} {\it For a given source $E$, find
the greatest lower bound $q_{\rm min}$ of all $q$'s with the following
property: For all $\epsilon >0$ there exists a coding/decoding
scheme based on $q$ qubits/signal with fidelity $1-\epsilon$.}\\
This problem may be considered either in the blind or visible
context, with the variation over encodings taken over the
appropriate class of maps in each case. Similarly, it may be
considered in the case of either of the fidelity criteria,
(LOCAL-FID) or (GLOBAL-FID).  We will say that the source $E$ can 
be coded (or compressed) at the rate $q_{\rm min}$.

Equivalently, the problem may be stated as follows: For a given 
source $E$, find $q_{\rm min}$ with the following property. Given
any $\delta >0$, (a)~if $q_{\rm min}+\delta$ qubits/signal are
available, then for every $\epsilon >0$ there exists a coding
scheme with fidelity $1-\epsilon$, and (b)~if $q_{\rm min}-\delta$
qubits/signal are available, then there exists an $\epsilon >0$
such that every coding scheme will have fidelity less than
$1-\epsilon$.

\section{Comparing the Formulation with Schumacher's\\
Coding for Pure States} 

The problem formulated above is intended to be a generalization of the 
scenario in Schumacher's theorem \cite{Schumacher95a,Jozsa94a} to the 
case of mixed input states.  Indeed, if the input states happen all to 
be pure states, then the above formulation reduces precisely to the 
situation of Schumacher's theorem. It is interesting to note that several 
of the distinctions made above collapse in the special case of {\em pure\/} 
input states.
\\[2mm] 
{\bf Proposition 4}. If the input states are all pure, then there is 
no distinction between the blind and visible problems.
\\[1mm] 
Proof: In Refs.~\cite{Jozsa94a,Barnum96b} an optimal coding/decoding scheme 
for the visible pure-state problem is described. This optimal scheme
turns out, remarkably, to be blind; i.e., knowledge of the
identities of the individual input signals gives Alice no further
benefit in the case of pure states. 
$\QED$

In \cite{Barnum96b} it is also shown that nonunitary decoding
operations are of no advantage (on the criterion (GLOBAL-FID)) in
decoding for the pure-state problem. In contrast, for mixed-state
signals, nonunitary decodings are generally essential for optimal
compression. This follows from Theorem~7 and \S 7 below.

Finally, the distinction between (LOCAL-FID) and (GLOBAL-FID) also
collapses for pure signal states.
\\[2mm] 
{\bf Proposition 5}. If the input states are all pure, then the two 
alternative fidelity criteria, (LOCAL-FID) and (GLOBAL-FID), become 
equivalent as $\epsilon$ is allowed to tend to zero.
\\[1mm] 
Idea of proof: We already know that the (GLOBAL-FID) criterion implies 
the (LOCAL-FID) criterion. Suppose that the (LOCAL-FID) criterion holds 
for a sequence of $\epsilon$ values tending to zero.  (Here we are
thinking of a sequence of coding/decoding schemes which all
operate within the resource constraint of $q$ qubits/signal where
$q>q_{\rm min}$.)  Then the reduced states $\tilde{\rho}_k$ of
$\tilde{\sigma}_N$ become arbitrarily close to the input states
$\rho_{i_k}$ which are {\em pure}. Hence $\tilde{\sigma}_N$ cannot
be much entangled since entanglement always shows up as {\em
impurity} in the reduced states $\tilde{\rho}_k$. Thus
$\tilde{\sigma}_N$ must approach the product state $\sigma_N$ and
(GLOBAL-FID) holds. 
$\QED$

As a consequence of Proposition 5, the awkward feature of (LOCAL-FID) 
described in Remark 2 does not arise in the coding of {\em pure\/} states.

\section{The $S(\bar{\rho})$ Upper Bound for $q_{\rm min}$}

Let $\bar{\rho}$ be the average density matrix of the input states:
\begin{equation}
\bar{\rho} = \sum_{i=1}^{n} p_i \rho_i\;.  \label{ave}
\end{equation}
\\[2mm]
{\bf Proposition 6}. $S(\bar{\rho})$ is an upper bound
for $q_{\rm min}$ under the criterion (GLOBAL-FID) (and hence also
under the criterion (LOCAL-FID)). 
\\[1mm]
Proof: For each $\rho_i$, choose a representative ensemble of pure states 
corresponding to $\rho_i$, so that we may view Alice as receiving an 
overall ensemble of pure states with density matrix $\bar{\rho}$.  By
Schumacher's theorem this may be transmitted to Bob with arbitrarily high 
fidelity by compressing to $S(\bar{\rho})$ qubits/signal.
$\QED$

Note that this compression preserves too much internal structure: Bob 
faithfully reconstructs Alice's chosen ensembles of pure states underlying 
the $\rho_i$'s rather than just the $\rho_i$'s themselves.  For our purposes 
it is sufficient for Bob to decode to any {\em other\/} representative 
ensemble for the $\rho_i$'s.  Hence we would expect that further compression 
is possible, and the examples in \S 7 below show that it generally is. 
Furthermore, the coding in Proposition 6 gives high fidelity relative to 
the stronger criterion (GLOBAL-FID); using the weaker (LOCAL-FID), one might 
expect even more compression.

In fact we can say more, embodied in the following theorem.
\\[2mm] 
{\bf Theorem~7} \cite{Jozsa94d} \cite{Lo95a}. For the stronger fidelity
criterion (GLOBAL-FID), if the decoding operation is required to be 
unitary (i.e., using only OP1 and OP2), then no further compression is 
possible, i.e., $q_{\rm min} = S( \bar{\rho})$.
\\[1mm] 
The proof is given in Appendix~A.

Note that for the pure-state coding theorem, the decoding may
indeed be taken to be unitary and (GLOBAL-FID) is used (being
equivalent to (LOCAL-FID) by Proposition 5), but we do not
necessarily wish to impose these conditions in the mixed-state
case.

\section{A Classical Analogue}

In the case of Schumacher's pure-state coding theorem, there is a
clear classical analogue, which has been well studied and completely 
solved, namely Shannon's noiseless coding theorem.  Though the classical 
analogue for the case of mixed states appears not to have been studied, 
it would involve the compression/communication of probability distributions.  
To formulate the classical problem, let there be a finite number of possible 
classical states, i.e., distinguishable alternatives (this is the analogue 
of our assumption of finite-dimensional Hilbert spaces), and identify the
input and output classical states with particular
orthonormal bases in input and output
Hilbert spaces.  Write probability weight functions on the sets of 
orthonormal pure states as column vectors ${\bf p} = (p_1,\ldots,p_n)^T$ 
of probabilities.  These classical probability distributions then 
correspond to commuting density operators diagonal in the input 
and output bases.

We may formulate classical preparation-blind coding or decoding procedures 
as multiplication of input probability vectors by a stochastic matrix 
${\bf A}$ (one with nonnegative entries whose columns sum to one): 
\beqa
{\bf p}_{\rm out} = {\bf A} {\bf p}_{\rm in}\;.
\eeqa
The stochasticity ensures that the matrix can be interpreted as a matrix 
of transition probabilities. As in the quantum case, preparation-visible
procedures are described by arbitary maps between the relevant spaces, 
in this case between the spaces of probability vectors.  The stochastic 
linear maps on the probability distributions correspond to a (convex) 
subset of the trace-preserving completely positive maps on density 
operators, and a given 
classical problem maps onto a corresponding quantum problem of sending
commuting density operators.  If we allow all possible trace-preserving 
completely positive maps, instead of just those which correspond to classical 
dynamics in the diagonalizing bases, we are using quantum means to deal 
with a classical problem, and we can compare the power of these quantum 
means to that of the purely classical means defined by restricting the 
allowable CP-maps to those that act as stochastic matrix multiplication 
in the given bases.


These notions and comparisons are illustrated in the following
examples, which are phrased in terms of the quantum language, i.e.,
viewing classical distributions as commuting mixed states.
\\[2mm]
{\bf Example 8}. We have two input states, $\rho_1 ,\rho_2 \in
{\cal H}_2$, 
\begin{equation}
\rho_1 \leftrightarrow  {\rm diag}(\alpha_1 , 1-\alpha_1 )\;,
\hspace{1cm}
\rho_2 \leftrightarrow  {\rm diag}(\alpha_2 ,1-\alpha_2 ) \;,
\end{equation}
which are simultaneously diagonal in a basis known to Alice and Bob. 
Let the prior probabilities for these two states be $p_1$ and $p_2$. 
Classically we may regard the two states as suitably biased coins, $C_1$ 
and $C_2$.  A preparer chooses a sequence of coins, $C_1$ or $C_2$ with 
probabilities $p_1$ and $p_2$, and tosses each of them once.  The sequence 
of outcomes is passed on to Alice.  Since Alice can look at the sequence of 
outcomes, we can regard the sequence of outcomes as the realization of ``Alice 
being given an unknown sequence of the two states.''   Notice that in the blind 
case, Alice cannot be given the actual coins that make up the input sequence,
for she could then toss each one many times and identify the coins in the 
sequence, which is impossible to do given a single instance of each quantum 
state in the sequence.  In contrast, in the visible case, Alice is given 
the sequence of coin names (or the actual coins, from which she could generate
the sequence of coin names), together with a sequence of outcomes.  In both 
cases, the objective of the protocol is to have Bob generate a sequence 
of outcomes that are governed by the same probabilities as Alice's input 
sequence of outcomes.  Thus we have the following classical problems.

{\bf Blind case}: A preparer chooses a sequence of coins, $C_1$ or $C_2$ 
with prior probabilities $p_1$ and $p_2$, tosses each of them a single time, 
and passes the sequence of outcomes on to Alice.  Alice ``codes'' her 
sequence of outcomes, and Bob ``decodes'' the result, obtaining an output 
sequence of outcomes.  The coding/decoding processes may involve probabilistic 
processes.  As before, Alice would like to compress the input sequence as 
much as possible for transmission.  A perfect coding/decoding scheme would
achieve the following: Suppose that in position 1 the preparer has used coin 
$C_2$; then, taking into account the probability of outcomes in tossing $C_2$ 
and all probabilistic processes involved in coding/decoding, the first entry 
in Bob's outcome sequence should have a probability distribution which is 
the {\em same\/} as for coin $C_2$.  A similar condition should apply at 
each position of the sequence.

This condition requires perfect fidelity of transmission of the 
distributions. In order to allow the usual situation of fidelity that 
approaches perfection only in an asymptotic limit of longer and longer 
block coding, we introduce a fidelity function for classical probability 
distributions. If ${\bf p} = (p_1,\ldots, p_n)^T$ and 
${\bf q} =(q_1,\ldots, q_n \}^T$ are two probability distributions 
on the same space, then the fidelity is defined by 
\begin{equation}
F_{\rm cl}({\bf p},{\bf q}) = 
\left(\sum_{i=1}^{n} \sqrt{p_i}\sqrt{q_i}\right)^2 \;,
\label{fcl}
\end{equation}
which is also known as the Bhattacharyya-Wootters distance (or overlap)
between the distributions. Notice that $F_{\rm cl}({\bf p},{\bf q}) =1$
iff ${\bf p}={\bf q}$.  The classical fidelity $F_{\rm cl}$ may be viewed as 
a special case of the Bures-Uhlmann fidelity (\ref{uhl}), i.e., 
$F_{\rm cl}({\bf p},{\bf q}) = F(\rho_1 , \rho_2 )$ for two commuting 
density operators, $\rho_1$ and $\rho_2$, that have $\bf p$ and $\bf q$ 
on their diagonals. 

The problem is then to find the minimum number of bits/signal which suffices 
to code the input string with asymptotically arbitrarily high fidelity. 
(A precise formulation is very similar to that given for the quantum problem 
in \S 2.)   There is an obvious upper bound on the minimum number of 
bits/signal: Alice may compress her outcome sequence to the Shannon entropy
of the average coin, $H(\bar{\alpha},1-\bar{\alpha})=S(\bar\rho)$ bits/signal, 
where $\bar{\alpha} = p_1 \alpha_1 + p_2 \alpha_2 $ is the average probability 
for the first outcome; Bob can decode the compressed sequence to produce an 
output outcome sequence that has asymptotically perfect fidelity.  Because 
we are dealing here with commuting density operators, this upper bound is 
the same as the $S(\bar\rho)$ upper bound of \S 5.

{\bf Visible case}: In this case Alice is fed the sequence of actual coin 
names, $C_1$ or $C_2$, in addition to outcomes of tossing each coin once. 
The blind-case upper bound of $H(\bar{\alpha},1-\bar{\alpha})=S(\bar\rho)$ 
bits/signal applies also to the visible case, but there is an additional
clear upper bound in the visible case: Alice may simply send Bob the full 
information of which coin to use at each stage; she can compress this data 
by the Shannon entropy of the prior distribution of coin choices, i.e.,
$H(p_1 ,p_2 )$ bits/signal.

Although we do not know the optimal number of bits/signal for this problem,
we now describe a purely classical coding/decoding scheme which beats
both bounds for some values of the parameters $p_1$, $p_2$, $\alpha_1$, and 
$\alpha_2$.
\\[2mm] 
{\bf Example 9}. Suppose that $\alpha_2\geq\alpha_1 $. Denote the coin toss 
outcomes by H and T, with H having probability $\alpha_i$ for coin $C_i$. 
Alice sends one of three possible messages, $M_0$, $M_1$, or $M_2$, to Bob
according to the following (probabilistic) coding scheme:
\begin{itemize}
\item{Regardless of the input coin ($C_1$ or $C_2$), Alice sends $M_0$
with probability $1-\alpha_2 +\alpha_1$.}
\item{If the message $M_0$ is {\em not\/} chosen (i.e. with probability 
$\alpha_2 - \alpha_1$), Alice sends $M_1$ if the coin is $C_1$ and $M_2$ 
if the coin is $C_2$.}
\end{itemize}
\noindent 
Bob responds to these signals as follows:
\begin{itemize}
\item{For $M_0$ Bob probabilistically generates H or T with
$\mbox{prob(H)}= \alpha_1 /(1-\alpha_2 +\alpha_1 )$ and
$\mbox{prob(T)}= (1-\alpha_2 )/(1-\alpha_2 +\alpha_1 )$.}
\item{
For $M_1$ Bob generates T with probability 1.
}
\item{
For $M_2$ Bob generates H with probability 1.
}
\end{itemize}
Curiously, in the latter two cases Bob actually learns the
identity of the coin yet he responds with a different
distribution! It is readily verified that for each position in the
sequence, taking into account the probabilistic choices in
coding/decoding, Bob's output result correctly represents the
result of one toss of the corresponding input coin.

The messages $M_0$, $M_1$, and $M_2$ are sent with probabilities
$1-\alpha_2 +\alpha_1$, $p_1 (\alpha_2 - \alpha_1 )$, and
$p_2 (\alpha_2 - \alpha_1 )$, so Alice can compress the sequence to
\[ \Xi = H\Bigl(  1-\alpha_2 +\alpha_1 , p_1 (\alpha_2 - \alpha_1 ),
p_2 (\alpha_2 - \alpha_1 )\Bigr) \mbox{ bits/coin toss.}  \]
If $p_1 = p_2 = \frac{1}{2}$ and $\alpha_1 \approx \alpha_2 \approx
\frac{1}{4}$, then $H(p_1 ,p_2 )=1$ and 
$S(\bar\rho)=H(\bar{\alpha},1-\bar{\alpha})\approx H(\frac{3}{4},\frac{1}{4})$, 
whereas $\Xi \approx 0$, thus beating the two bounds. (For some other values 
of the parameters $\Xi$ exceeds both bounds).

It has been conjectured that the minimum number of bits/signal for
this classical problem (and its natural generalization too many
classical distributions) should be the {\em mutual information\/}
$H(\bar{\alpha}, 1-\bar{\alpha})-p_1 H(\alpha_1 ,1-\alpha_1 )
-p_2 H(\alpha_2 ,1-\alpha_2 )$, even if global fidelity is required,
but this has resisted proof/disproof so far. (This would coincide with 
the lower bound given in $\S$8.)  In Example 12 below we will describe a 
quantum protocol for this problem which is better than all the above 
protocols.

\section{Examples of Compression beyond $S(\bar{\rho})$}

We now return to our main question of quantum coding for general sources 
of mixed states.  Though the problem of the optimal value of $q_{\rm min}$ 
remains unsolved, we describe here a series of interesting examples of 
compression beyond the $S(\bar\rho)$ upper bound given in \S 5.  These 
examples reveal something of the intricacy of this problem.  (Notice that 
Example~9 already provides a case of compression beyond the $S(\bar\rho)$
bound in the classical context.)  In the next section we will derive a 
lower bound for $q_{\rm min}$.
\\[2mm] 
{\bf Example 10} (Trivial cases).  The following two situations are blind, 
but Alice may reliably identify the input states, thus making them visible.
\\[1mm] 
(a) Suppose that there is only one possible input signal 
$\rho =\frac{1}{2}I$ so $\bar{\rho} = \frac{1}{2}I$ and 
$S(\bar{\rho})= 1$ qubit/signal.  Yet Alice need not send anything 
at all; i.e., we may compress to 0 qubits/signal.
\\[1mm] 
(b) Suppose that the input signals $\rho_i$ with prior probabilities 
$p_i$ are supported on {\em orthogonal\/} subspaces. (The support of a 
mixed state is defined as the subspace spanned by all eigenvectors 
belonging to {\em nonzero\/} eigenvalues.) Thus Alice may reliably 
measure the identity of the inputs and compress the resulting data to 
$H(p_1,\ldots, p_n)$ qubits/signal. Now for orthogonally supported states
we have generally
\begin{equation}
S(\bar{\rho}) = H(p_1 ,\ldots, p_n ) + \sum p_i S(\rho_i )
\geq H(p_1 ,\ldots, p_n )\;. 
\end{equation}
$\QED$
\\[2mm]
{\bf Example 11} (A nontrivial blind example with noncommuting mixed 
input states).  There are two signal states, $\rho_1$ and $\rho_2$, in 
$m+n$ dimensions with prior probabilities $p_1$ and $p_2$.  The states have 
a block-diagonal form,
\[ \rho_1 = \mbox{diag}\Bigl(\epsilon \sigma_1 , (1-\epsilon ) \tau_1\Bigr)\;,
\qquad
\rho_2 = \mbox{diag}\Bigl(\epsilon \sigma_2 , (1-\epsilon ) \tau_2\Bigr)\;, \]
where $\sigma_1$ and $\sigma_2$ are density matrices of size $m\times m$
and $\tau_1$ and $\tau_2$ are density matrices of size $n\times n$.  Writing
\[ \bar{\rho} = p_1 \rho_1 + p_2 \rho_2\;,\qquad 
\bar{\sigma} = p_1 \sigma_1 + p_2 \sigma_2\;,\qquad
\bar{\tau} = p_1 \tau_1 + p_2 \tau_2\;, \]
ones easily sees that 
\begin{equation}
S(\bar{\rho})= H(\epsilon , 1-\epsilon ) +\epsilon S(\bar{\sigma})
+(1-\epsilon) S(\bar{\tau})\;.  
\label{sp}
\end{equation}
In the $S(\bar{\rho} )$ coding scheme of Proposition 6, we may interpret this
formula as follows. For a sequence of inputs Alice first measures the 
$\sigma$-space versus the $\tau$-space --- projecting the input state into 
whichever space is the outcome--- and she compresses the resulting string 
of subspace names to $H(\epsilon ,1-\epsilon )$ bits/name.  If the outcome space 
was ``$\sigma$-subspace,'' a result that occurs a fraction $\epsilon$ of 
the time, she compresses the post-measurement state to $S(\bar{\sigma})$ 
qubits/signal, and similarly if the outcome was ``$\tau$-subspace,'' 
which occurs $(1-\epsilon)$ of the time, she compresses to $S(\bar{\tau})$ 
qubits/signal.  Thus the total sending resources is the sum of these three 
terms in (\ref{sp}).

Now suppose that $\sigma_1 \neq \sigma_2$, but that $ \tau_1 = \tau_2
\equiv \tau$. Then, in the case that Alice's measurement outcome is
``$\tau$-subspace,'' a result that also becomes known to Bob through the
communication of the subspace names, she need not send the post-measurement 
state at all, as Bob already knows it (i.e. $\tau$) and can construct it 
himself.  Thus we may drop the last term in (\ref{sp}) and communicate the 
mixed states (with perfect fidelity) using only 
$H(\epsilon , 1-\epsilon ) + \epsilon S(\bar{\sigma})$ qubits/signal, 
which is less than $S(\bar{\rho})$ by an amount $(1-\epsilon)S(\bar{\tau})$.
$\QED$
\\[2mm]

{\bf Example 12} (Visible coding by purification of the input
states).  The general idea here (cf.~also \cite{Horodecki98a}) is that
in the visible situation, Alice may build purifications of the input mixed 
states and send these purifications (which are {\em pure\/} states) to Bob 
utilizing the compression of Schumacher's pure state coding theorem. On
reception Bob regains the mixed states by selecting a suitable subsystem 
of each decoded purification state.  

As a first example, consider a special case of states of the form in Example~8.  
There are two possible input states,
\begin{equation}
\rho_1 = \mbox{diag}(\epsilon , 1-\epsilon )\;,
\qquad\rho_2 = \mbox{diag}(1-\epsilon ,\epsilon )\;,
\label{in}
\end{equation}
with equal prior probabilities $p_1 = p_2 = \frac{1}{2}$.
Hence $S(\bar{\rho})=1$ and $H(p_1 ,p_2 )=1$.  After constructing
purifications $\ket{\psi_1}$ and $\ket{\psi_2}$, Alice's task is
to send a 50/50 mixture of $\ket{\psi_1}$ and $\ket{\psi_2}$.
Thus to get the greatest benefit from Schumacher compression, the
purifications should be chosen so that their ensemble has least
von Neumann entropy; i.e., the two purifications should be as
parallel as possible. According to Bures and Uhlmann's basic theorem
\cite{Bures69a,Uhlmann76a,Jozsa94b}, 
the minimum possible angle $\theta_{\rm min}$
between purifications of $\rho_1$ and $\rho_2$ is given by
\[ \cos^2\!\theta_{\rm min} = F(\rho_1 , \rho_2 )\;. \] \label{BUlimit}
Moreover, a 50/50 mixture of states at angle $\theta_{\rm min}$ has 
entropy 
\[ S_{\rm min} = 
H\!\left(\frac{1+\cos \theta_{\rm min}}{2},
\frac{1-\cos \theta_{\rm min}}{2}\right)\;,
\]
which gives the Schumacher limit of qubits/signal in compression
by this method.

For the states in (\ref{in}) we readily compute
\[  F(\rho_1 ,\rho_2 ) = 4\epsilon (1-\epsilon )\;, \]
so that Alice may compress the purification ensemble to
\[ \Upsilon (\epsilon ) = H\!\left( \frac{1}{2}+\sqrt{\epsilon (1-\epsilon)},
\frac{1}{2}-\sqrt{\epsilon (1-\epsilon)}\right) \quad
\mbox{qubits/signal}\;,   \] 
which is better than $S(\bar{\rho})$ or $H(p_1 ,p_2 )$, being equal to these 
only when $\epsilon$ is 0 or 1.

Note that the purely classical compression method of Example 9 applies to 
this case, too.  The relevant parameters are $p_1 = p_2 = \frac{1}{2}$, 
$\alpha_1 = \epsilon$, $\alpha_2 = 1-\epsilon$, and 
$0\leq \epsilon \leq \frac{1}{2}$.  The method of Example 9 gives compression 
to 
\[ \Xi (\epsilon) = 
H\!\left(2\epsilon , \frac{1}{2}-\epsilon , \frac{1}{2} -\epsilon\right)
=H(2\epsilon,1-2\epsilon)+1-2\epsilon\quad
\mbox{qubits/signal}\;,   \] 
and we get
\[ \Xi (\epsilon ) \geq \Upsilon (\epsilon ) \qquad\mbox{for}
\quad 0\leq \epsilon \leq \frac{1}{2}\;,  \] 
with equality only for $\epsilon$ is 0 or 1/2.  Thus, whenever the states
$\rho_1$ and $\rho_2$ of Eq.~(\ref{in}) are mixed, the quantum purification 
compression beats the classical method of Example~9. 
$\QED$
\\[2mm] 
{\bf Remark 13} (A simple construction of optimally parallel purifications 
for commuting states).  Given any mixed state in diagonal form 
$\rho = \mbox{diag}(p_1 , \ldots, p_n )$, we may immediately write down 
a canonical purification:
\[ \ket{\psi} = \sum_{i=1}^{n} \sqrt{p_i} \ket{e_i}\otimes \ket{e_i}\;, \]
where $\{ \ket{e_k} \}$ is the diagonalizing basis for $\rho$.  Given two 
such states,
\[ \rho_1 = \mbox{diag}(p_1 ,\ldots, p_n )\;,\qquad
\rho_2 = \mbox{diag}(q_1 ,\ldots, q_n )\;, \]
the canonical purifications clearly satisfy
\[ |\bra{\psi_1}  \psi_2 \rangle |^2 =
\left(\sum_{i=1}^{n} \sqrt{p_i}\sqrt{q_i} \right)^2 = F(\rho_1 ,\rho_2 ) =
\mbox{(Bures-Uhlmann limit for $\cos^2\!\theta$).}  \] 
Thus for simultaneously diagonal states, the canonical purifications are
always optimally parallel.

Notice that the diagonal entries of $\rho_1$ and $\rho_2$ are classical 
distributions $\bf p$ and $\bf q$ and that
\[  |\bra{\psi_1}  \psi_2 \rangle |^2 = F(\rho_1 , \rho_2 ) =
F_{\rm cl}({\bf p},{\bf q})\;.  \]
The construction of optimally parallel purifications converts the
Bhattacharyya-Wootters overlap of classical distributions into quantum 
overlap of pure quantum states. In this way the methods of {\em quantum\/}
coding may be applied to problems of compression of {\em classical\/} 
probability distributions.

Suppose now that we have two or {\em more\/} simultaneously diagonal states,
\[ \rho^{(a)} = \mbox{diag}(p_{1}^{(a)}, \ldots, p_{n}^{(a)} )\;,\qquad
a=1,\ldots, K \;. \] 
Then their canonical purifications $\ket{\psi^{(a)}}$ have the remarkable 
property that they are all simultaneously pairwise maximally parallel.  
Recall that Uhlmann's theorem gives a limit on how parallel purifications 
can get for any {\em pair\/} of mixed states. It does not follow that this
optimal parallelness can be {\em simultaneously\/} achieved by purifications 
of three or more states.  
Yet for simultaneously diagonal states, this optimal simultaneous parallelism 
is achieved by the canonical purifications.

It seems unlikely, however, that maximum parallelness gives the best set 
of purifications for the purpose of mixed-state compression when there are 
three or more signal states.  Jozsa and Schlienz \cite{Jozsa99b} have shown 
the existence of pairs of pure-state ensembles $\{p_i, \ket{\psi_i}\}$ and 
$\{p_i, \ket{\chi_i}\}$ for which all homologous pairs in the second ensemble are 
less parallel (i.e., 
$\forall i,j~ |\inner{\chi_i}{\chi_j}| \ge |\inner{\psi_i}{\psi_j}|$), but 
for which the entropy of the second ensemble is nevertheless smaller.  This 
phenomenon is expected to persist under the added constraint that the states
involved are purifications of the given mixed states.
$\QED$
\\[2mm]
{\bf Remark 14}. If Alice sends Bob the canonical purification of
$\rho$,
\[
\ket{\psi} = \sum \sqrt{p_i} \ket{e_i}\otimes \ket{e_i}\;,
\]
she is actually supplying him with  two copies of $\rho$ -- one for each 
of the two subsystems of the purification.  Therefore one suspects that 
this compression is not optimal, at least when the criterion (LOCAL-FID) 
is used.  To benefit from this observation, we might try to construct 
purifications each of which codes two signal states, one in each subsystem
of the purification.  To do this, the two signal states must have the
same eigenvalues, but they need not be identical (e.g., as occurs in 
Example~15 below).  Thus the signal states would purify each other in 
pairs at the expense of introducing strong entanglement in the output 
signal sequence.  This construction would have high (LOCAL-FID) fidelity, 
but low fidelity for the (GLOBAL-FID) criterion.  Of course, even with the 
stronger criterion (GLOBAL-FID), it is not clear that the compression of
Example 12 is optimal.
\\[2mm]
{\bf Example 15}. (The ``photographic negative'' example, another application 
of compression by purification).  Suppose that we have $d$ possible input 
signals $\rho_i$, where $\rho_i$ is the $d\times d$ diagonal density matrix 
with equal entries $1/(d-1)$ along the diagonal except for the $i$th entry
which is zero:
\[ \rho_i = \frac{1}{d-1} \mbox{diag}(1,1,\ldots,1,0,1,\ldots,1)\;,
\qquad\mbox{where $0$ is in the $i$th place.}  \]
The signals all have equal prior probabilities $p_i = 1/d$, giving
$\bar\rho=I/d$.

The canonical purifications in ${\cal H}_d \otimes {\cal H}_d$
all lie in a $d$-dimensional subspace spanned by $\{\ket{e_i}\otimes
\ket{e_i}\}$, where $\{\ket{e_i}\}$ is the diagonalizing basis of
the $\rho_i$'s.  A direct calculation shows that the equally weighted 
mixture of purifications in this $d$-dimensional subspace is a density 
matrix
\[ \rho = \frac{d-2}{d-1}\ket{\psi}\bra{\psi} + \frac{1}{d-1}
\,\frac{I}{d}\;,  \]
where $I$ is the identity matrix in the $d$-dimensional subspace, and
\[ \ket{\psi} = \frac{1}{\sqrt{d}}\sum_{i=1}^d\ket{e_i}\otimes\ket{e_i} \]
is a maximally entangled state.  Thus $\rho$ can be viewed as a mixture of 
a totally mixed state $I/d$, with probability $1/(d-1)$, and a maximally 
entangled pure state $\ket{\psi}\bra{\psi}$, with probability $(d-2)/(d-1)$.  
Changing to a basis in which $\ket{\psi}$ is the first basis vector, we can 
easily determine the eigenvalues of $\rho$ to be a nondegenerate eigenvalue
$(d-1)/d$ and $(d-1)$ degenerate eigenvalues $1/d(d-1)$. A short calculation 
gives
\[ q = S(\rho) = 
H\!\left(\frac{d-1}{d},\frac{1}{d}\right)+\frac{1}{d}\log(d-1)
=\frac{2}{d}\log (d-1) - \log\!\left (1-\frac{1}{d}\right)
\quad\mbox{qubits/signal} \]
for the compression scheme.  Note that $q \rightarrow 0$ as 
$d \rightarrow \infty$.  

Introducing the Holevo quantity for the ensemble $E=\{p_i,\rho_i\}$, 
\[ \chi(E) = S(\bar{\rho}) - \sum p_i S(\rho_i ) =  
- \log\!\left (1-\frac{1}{d}\right) \;, \]
we find
\[ q = \chi +\frac{2}{d}\log (d-1) \;,  \]
so $q\rightarrow \chi$ as $d\rightarrow \infty$.  Note that although we 
described this construction in terms of block coding the ensemble of 
canonical purifications for all the signals, it also provides canonical 
purifications for the ensemble of $N$-block mixed states.  Nonetheless, 
for finite $d$, the above bound remains greater than the Holevo bound.  
Thus, if the conjecture that the Holevo bound is achievable by visible 
compression is correct, then, perhaps surprisingly, canonical purification 
is a suboptimal method of compression.
$\QED$

\section{A lower bound on the rate of mixed-state compression}

There is a simple argument that the Holevo quantity for an ensemble
$E=\{p_i,\rho_i\}$ of mixed states is a lower bound on the rate at which such 
an ensemble can be coded.  Here we use the global fidelity criterion
(GLOBAL-FID), and encoding may be blind or visible.
This argument uses the result, shown for 
pure-state ensembles by Hausladen, Jozsa, Schumacher, Westmoreland, and 
Wootters \cite{Hausladen96a} and for general mixed-state ensembles by 
Holevo \cite{Holevo98a} and by Schumacher and Westmoreland \cite{Schumacher97c}, 
that the Holevo quantity for an ensemble $E$ is the capacity for classical 
information transmission using the states in the ensemble $E$ as an alphabet.  
The gist of the argument is that if an ensemble of mixed states could be 
coded at a rate lower than its Holevo quantity, even with preparation-visible 
encoding, then one could code a Holevo quantity's worth of classical information 
into those mixed states, compress them to an ensemble on a channel space 
of size smaller than the Holevo quantity (per use), recover the original 
ensemble with high fidelity, and therefore recover the classical information.
But since the classical information capacity of an ensemble of states cannot 
be larger than the log of the dimension of its Hilbert space (since this 
is greater than or equal to $\chi$ for any ensemble), this is impossible.

To formalize this argument, 
consider an ensemble (or source) $E=\{p_i,\rho_i\}$ of mixed states
$\rho_i$ with probabilities $p_i$ on a Hilbert space $H_d$ of dimension $d$.  
The Holevo quantity for this ensemble is
\beqa \chi(E) \equiv
S\Biggl(\sum_i p_i \rho_i\Biggr) - \sum_i p_i S(\rho_i) \;.
\eeqa 
A sequence of $N$ signals from this source gives a state drawn from the 
ensemble 
\beqa E\on = \{ p_{i_1} p_{i_2} \cdots p_{i_N} ,
\rho_{i_1}\otimes \rho_{i_2} \otimes \cdots \otimes \rho_{i_N}\}
\;. 
\eeqa 
We introduce the notation $f(A) = \{p_i, f(\rho_i)\}$ for the ensemble 
obtained by applying a map $f$ to the states of the ensemble $A$, and 
we write $B({\cal H})$ for the space of bounded operators on a Hilbert 
space $\cal H$.

For two ensembles with the same probabilities, $A = \{p_i,\rho_i\}$ and 
$B = \{p_i, \sigma_i\}$, we define an average fidelity by
\beqa \overline{F}(A,B) = \sum_i p_i
F(\rho_i, \sigma_i)\;. 
\eeqa 
In proving the main theorem of this section, we will need a lemma 
that bounds the absolute value of the difference in the Holevo quantities 
for two ensembles in terms of their average fidelity, provided
the average fidelity is high enough.
\\[2mm]
{\bf Lemma 16}. If $\overline{F}(A,B) > \sqrt{35/36}$, then 
\beqa
\label{eqtn: basic2} 
|\chi(A) - \chi(B)| \le (2 + 2\sqrt{2})
\sqrt{1 - \overline{F}(A,B)}\log{d} + 1 \;, 
\eeqa 
where $d$ is the dimension of the state space of $A$ and $B$.
\\[1mm]
The proof is given in Appendix~\ref{holevocontinuity}.

Our formulation of the mixed-state compression problem for the fidelity
criterion (GLOBAL-FID) can now be stated succinctly.  Relative to
(GLOBAL-FID), the source $E$ can be coded (or compressed) at a rate 
$q$ if there exists a channel Hilbert space $C$ with $q = \log({\rm dim}\,C)$ 
and encoding/decoding schemes $\{e\pn,\cd\pn\}$,  
\beqa 
e\pn:\;B(H_d\on) \rightarrow B(C\on) \;, \qquad
\cd\pn:\;B(C\on) \rightarrow B(H_d\on) \;, 
\eeqa 
such that
\beqa \label{asyaveragefid} 
\lim_{N \rightarrow \infty} \overline F(E\on,F\pn) = 1 \qquad
\mbox{(GLOBAL-FID)}\;, 
\eeqa 
where 
\beqa 
F\pn\equiv\cd\pn\circ e\pn(E\on) 
\eeqa 
is the ensemble after decoding.  We require that the encodings $e\pn$ 
take density operators to density operators and that the decodings 
$\cd\pn$ be trace-preserving completely positive linear maps.  Permitting 
the encodings to be arbitrary maps on density operators allows for 
preparation-visible encoding; if $e\pn$ is a trace-preserving completely 
positive linear map $\ce\pn$, then the compression is preparation-blind.

The argument outlined at the beginning of 
this section can be formalized in the
following theorem.
\\[2mm]
{\bf Theorem~17.} For the fidelity criterion (GLOBAL-FID) and for both 
blind and visible encodings, the Holevo quantity $\chi(E)$ for an ensemble 
$E=\{p_i,\rho_i\}$ is a lower bound for $q_{\rm min}$.
\\[1mm]
{\it Proof:} Suppose that the ensemble $E = \{ p_i, \rho_i\}$ can be 
compressed at a rate $q < \chi(E)$ with asymptotically high fidelity
(Eq.~(\ref{asyaveragefid})), whether preparation-blind or preparation-visible.  
Consider the ensemble of channel states (density matrices)
\beqa
W\pn= \{ p_{i_1} p_{i_2} \cdots p_{i_N}, w_{i_1i_2\ldots i_N} \}\;,
\eeqa
where 
\[ w_{i_1i_2\ldots i_n}\equiv e\pn(
\rho_{i_1}\otimes \rho_{i_2} \otimes \cdots \otimes \rho_{i_N}) \]
is the encoded state corresponding to the unencoded source state
$\rho_{i_1}\otimes \rho_{i_2} \otimes \cdots \rho_{i_N}$.  The Holevo
quantity for $W\pn$ satisfies
\beqa
\label{eqtn: basic}
\chi(W\pn)\le S\!
\pmatrix{\mbox{average density}\cr\mbox{operator of $W\pn$}}\le Nq < 
N\chi(E)\;,
\eeqa
where $Nq$ is the the log of the dimension of the channel Hilbert space 
for $N$ blocks of channel.  

Consider now the following procedure for using the $N$-block operators 
$w_{i_1\ldots i_N}$ as an alphabet to send classical information.  Make 
codewords out of strings of $M$ of these operators.  Prune them as one 
would if one were coding using the operators $\rho_{i_1\ldots i_N}$ of 
the ensemble $F\pn$ in the Holevo/Schumacher/Westmoreland procedure for 
attaining $\chi(F\pn)$ as classical capacity.  As the first step in the
decoding procedure, convert them using the decoding $(\cd\pn)^{\otimes M}$
into strings of the operators $\rho_{i_1\ldots i_N}$ of the ensemble $F\pn$.  
Then apply the decoding measurement appropriate to that ensemble.  

This procedure clearly uses the $N$-block ensemble $W\pn$ to transmit 
classical information at the rate $\chi(F\pn)$ per $N$ blocks. But by 
assumption [cf.~(\ref{asyaveragefid})], the ensemble $F\pn$ has, at large 
enough $N$, arbitrarily high fidelity to the original ensemble $E\on$.  
Hence, applying Lemma~16 to the ensembles $E\on$ and $F\pn,$ whose states 
lie in $d^N$-dimensional Hilbert spaces, one finds 
\beqa \left|\,\chi(E) -
\frac{\chi(F\pn)}{N}\,\right| \le (2 + \sqrt{2}) 
\sqrt{1 - \overline{F}(E\on,F\pn)}\log{d} + \frac{1}{N} \;. 
\eeqa
Thus for large enough $N$, $\chi(F\pn)/N$ is arbitrarily
close to $\chi(E)$, which is greater than $\chi(W\pn)/N$ by at least
an amount $\chi(E)-q$, independent of $N$.  So for large enough $N$, 
$\chi(F\pn)$ exceeds $\chi(W\pn)$, contradicting the fact that the 
classical capacity of the ensemble $W\pn$ is $\chi(W\pn)$.  We conclude 
that the compression rate $q$ must satisfy $q\ge\chi(E)$.
$\QED$

M. Horodecki \cite{Horodecki98a} has independently derived the lower bound
of Theorem 17, 
using the nonincrease of the Holevo quantity under completely positive maps. 
This nonincrease is an easy consequence of the 
monotonicity of relative entropy under such maps \cite{Lindblad73a,
Uhlmann77a}, 
and therefore of Lieb's fundamental concavity theorem \cite{Lieb73b}). 
(A good treatment of all of these is to be found in \cite{Ahlswede99a}.)

A special case of Theorem 17 is the lower bound of $S(\overline{\rho})$
qubits per source signal on the rate of compression of ensembles of pure 
states.  This lower bound was established for preparation-blind encodings 
and unitary decodings in \cite{Schumacher95a}; 
for arbitrary (preparation-blind 
or preparation-visible) encodings and unitary decodings in \cite{Jozsa94a}; 
and, by somewhat technical arguments, for arbitrary encodings and decodings 
using completely positive trace-preserving maps in \cite{Barnum96a}.  The 
present result allows for arbitrary encodings and decodings using completely
positive trace-preserving maps, so it provides an alternative and perhaps 
more satisfying derivation of the most general form of the pure-state lower 
bound.

The lower bound in Theorem 17 
raises
the fundamental open question of whether the bound is achievable 
(with global fidelity) with either blind or visible encoding.  If not,
one would like an expression for the achievable rate in both cases.
Even for transmitting classical mixed states, the question of the best
achievable
rate remains open, in both the variant allowing quantum means of 
compression and that requiring only classical means.

\section*{Acknowledgments}
Much of the work reported here was 
carried out at the Center for Advanced Studies of 
the University of New Mexico in August 1995.
This work was supported in part by Office of Naval Research Grant
No.\ N00014-93-1-0116, National Science Foundation Grant
No.~PHY-9722614, the Institute for Scientific Interchange
Foundation, Turin, Italy and Elsag, a Finmeccanica company. 
RJ is supported by the U.K.  Engineering and 
Physical Sciences Research Council. RJ and HB are supported by
the European Union project QAIP, IST-1999-11234.
\begin{appendix} \label{unitarycoding}

\section{Proof of Theorem~7}

Proposition~6 may be used for part of the proof, but we give a different argument 
that utilizes properties of the Bures-Uhlmann fidelity function throughout. 
We first establish two lemmas which are direct Bures-Uhlmann fidelity analogues 
of Lemmas 1 and 2  in \cite{Jozsa94a}.
\\[2mm] 
{\bf Lemma A1}: Let $\rho$ and $\rhoprime$ be mixed states on ${\cal H}_n$
with $\rhoprime$ supported on a $d$-dimensional subspace $D$.  Then 
$F(\rho , \rhoprime )$ is less than the sum of the $d$ largest eigenvalues
of $\rho$, which we write as $1-\eta$.
\\[1mm]
{\em Proof of Lemma A1:} We use the fact that
\begin{equation}
F(\rho ,\rhoprime) = 
\mbox{inf trace }\rho A \mbox{ trace }\rhoprime A^{-1} \;,
\end{equation}
where the infimum is over all strictly positive operators $A$ \cite{Alberti83a}.  
Choose
\[ A= \left\{ \begin{array}{ll} 
       I & \mbox{on $D$,} \\
       \epsilon I  &  \mbox{on $D^{\perp}$ (for any $\epsilon > 0$).}
	      \end{array}
\right. \]
Then
\[ \mbox{trace } \rhoprime A^{-1} = 1 \;, \]
and
\[ \mbox{trace } \rho A = \mbox{trace }\rho D + \epsilon \, \mbox{trace }
\rho D^{\perp} \leq 1-\eta + \epsilon \, \mbox{trace } \rho D^{\perp}
\leq 1-\eta \;. \]
Hence
\[ F(\rho ,\rhoprime) \leq \mbox{trace }\rho A \mbox{ trace }\rhoprime A^{-1}
\leq 1-\eta \;, \]
as required.
$\QED$

To set the stage for the second lemma, consider a density operator $\rho$ on 
${\cal H}_n$.  Denote the eigenvalues of $\rho$ in decreasing order by
$\lambda_i$, $i=1,\ldots,n$.  Let $D$ be the $d$-dimensional subspace spanned 
by the eigenvectors belonging to the $d$ largest eigenvalues of $\rho\,$; 
denote the sum of these $d$ eigenvalues by $1-\eta$.   Denote the projector 
onto $D$ by $\Pi$, and let $\ket{0}$ be any pure state in $D$.  Now consider 
the density operator 
\[ \rhoprime \equiv \Pi \rho \Pi + \eta \proj{0} \;. \]
This density operator can be obtained from $\rho$ by first applying the binary 
measurement that projects onto $D$ (outcome ``1'') or onto $D^\perp$ 
(outcome ``0'') and then, if the outcome is 0, substituting $\proj{0}$ 
for the post-measurement state.  With these preliminaries, the lemma can be 
stated as follows.
\\[2mm]  
{\bf Lemma A2}: $F(\rho,\rhoprime)\ge1-2\eta$.
\\[1mm]
{\em Proof of Lemma A2:} If we write $\rho$ in its orthonormal eigenbasis, 
\[ \rho = \sum_{i=1}^{n} \lambda_i \proj{e_i} \;, \]
$\rhoprime$ becomes
\[ \rhoprime = 
\sum_{i=1}^{d} \lambda_i \proj{e_i}+\eta \proj{0} \;. \]
Introduce the following purifications of $\rho$ and $\rhoprime$:
\[ \ket{\phi} = \sum_{i=1}^{n} \sqrt{\lambda_i} \ket{e_i}\otimes \ket{f_i} \;, \]
\[ \ket{\phi '}= \sum_{i=1}^{d} \sqrt{\lambda_i} \ket{e_i}\otimes\ket{f_i}
+ \sqrt{\eta} \ket{0}\otimes\ket{g} \;. \]
Here the vectors $\ket{f_i}$, $i=1,\ldots,d$ are orthonormal, and $\ket{g}$ 
is orthogonal to each $\ket{f_i}$.  Since fidelity is the maximum absolute 
value of the inner product of purifications, we have
\[ F(\rho ,\rhoprime) \geq |\langle\phi|\phi'\rangle|^2
=\left(\sum_{i=1}^{d} \lambda_i\right)^{\!2}=(1-\eta )^2 \geq 1-2\eta \;, \] 
as required.
$\QED$
\\[2mm] 
{\em Proof of Theorem~7:\/} Suppose that we compress to $S(\bar{\rho})-\delta$
qubits/signal by any coding method whatsoever.  Then if the decoding scheme 
is unitary, the decoded state $\tilde{\sigma}_N$ of an input string 
$\sigma_N$ of length $N$ is supported in $N(S(\bar{\rho})-\delta)$ qubits. 
Yet the density matrix for strings of length $N$ is $\bar{\rho}\on$, and by 
a standard typical sequences result (cf.~\cite{Jozsa94a}), the sum of the
$2^{N(S(\bar{\rho})-\delta )}$ largest eigenvalues of $\bar{\rho}\on$ becomes 
arbitrarily small with increasing $N$.  Hence, by Lemma A1, 
$F(\sigma_N , \tilde{\sigma}_N ) $ is arbitarily small, too, and the 
fidelity cannot be high by the (GLOBAL-FID) criterion.
$\QED$

On the other hand, if $S(\bar{\rho})+\delta$ qubits/signal are available,
then Lemma A2 provides an explicit high-fidelity coding scheme,
with $D$ being the $2^{N(S(\bar{\rho})+\delta )}$-dimensional subspace
spanned by the $2^{N(S(\bar{\rho})+\delta )}$ weightiest eigenvectors of 
$\bar{\rho}\on$.  

\section{Proof of Lemma 16} \label{holevocontinuity}

The proof uses the following inequality (proved in \cite{Barnum2000a}): 
\beqa \label{eqtn: lemma} 
|S(\rho_1) - S(\rho_2)| \le 2 \sqrt{1 - F(\rho_1, \rho_2)} \log{d} + 1 \;,
\eeqa 
which is valid if 
\beqa \label{eqtn: validity condition} 
2 \sqrt{1 - F(\rho_1, \rho_2)} < \frac{1}{3}\;. 
\eeqa 
We use this to obtain a similar relation, but with the average ensemble 
fidelity in place of the fidelity on the right hand side. Let 
$A = \{p_i,\rho_i\}$ and $B = \{p_i,\sigma_i\}$ again denote two 
mixed-state ensembles having the same probabilities.  Letting 
$\rho \equiv \sum_i p_i \rho_i$ and  $\sigma \equiv \sum_i p_i \sigma_i$, 
we have an inequality involving the error measure $\sqrt{1 - F(\rho,\sigma)}$.  
We need to convert this into one involving the error measure 
$1 - \overline{F}(A,B)$.  Defining yet another error measure 
$\delta \equiv 1 - \sqrt{F(\rho,\sigma)}$, simple algebra gives 
$\sqrt{1-F(\rho,\sigma)} = \sqrt{\delta(2 - \delta)}$.  The double concavity 
of $G(\rho_1, \rho_2) \equiv \sqrt{F(\rho_1, \rho_2)}$ (proved in 
Appendix~\ref{appendix: double concavity}) gives 
\beqa 
\sum_i p_i G(\rho_i, \sigma_i) \le  
G\Bigl(\sum_i p_i \rho_i, \sum_i p_i \sigma_i\Bigr) 
= G(\rho, \sigma)\;. 
\eeqa
Hence 
\beqa \delta \equiv 1 - G(\rho, \sigma) &\le& 1 - \sum_i p_i
G(\rho_i, \sigma_i) \nonumber \\ 
&\le& 1 - \sum_i p_i F(\rho_i, \sigma_i) = 
1- \overline{F}(A,B) \equiv \epsilon \;. 
\eeqa
Therefore, we have the inequality 
\beqa \label{eqtn: ensemblecontinuity} 
|S(\rho) - S(\sigma)| & \le & 2
\sqrt{\delta(2-\delta)} \log d + 1 \nonumber \\ 
& \le & 2
\sqrt{\epsilon(2-\epsilon)} \log d + 1 \nonumber \\ 
& \le & 2
\sqrt{2} \sqrt{1 - \overline{F}(A,B)} \log{d} + 1\;. 
\eeqa 
This inequality is valid provided (\ref{eqtn: validity condition})
holds, which is certainly true if 
\beqa \label{eqtn: validitycondition2} 
0 \le \epsilon < 1 - \sqrt{35/36} 
\qquad\Longleftrightarrow\qquad
\overline F(A,B) > \sqrt{35/36} \;.
\eeqa

Furthermore, we can also use the inequality~(\ref{eqtn: lemma}) to
bound the difference in the average entropies for the two ensembles 
of $d$-dimensional states, 
\beqa\label{another equation}
\Bigl|\,\sum_i p_i S(\rho_i)\!\!&-&\!\!\sum_i p_i
S(\sigma_i)\Bigr| \nonumber \\ 
&\le& \sum_i p_i |S(\rho_i) -
S(\sigma_i)| \nonumber \\ 
&\le& \sum_i  p_i\!\left(2\sqrt{1 -
F(\rho_i,\sigma_i)}\log{d} + 1\right) \nonumber \\ 
&\le& 2 \sqrt{1
- \sum_i p_i F(\rho_i,\sigma_i)} \log{d} + 1 \nonumber \\
&\equiv& 2 \sqrt{1 - \overline{F}(A,B)} \log{d} + 1 \;. 
\eeqa
Combining Eqs.~(\ref{eqtn: ensemblecontinuity}) and (\ref{another equation})
yields the desired result~(\ref{eqtn: basic2}). 
$\QED$ 

\section{Double concavity of $G(\rho_1,\rho_2)$}
\label{appendix: double concavity} 

In this Appendix we show that 
\begin{equation}
G(\rho_1,\rho_2)\equiv\sqrt{F(\rho_1,\rho_2)}
=\mbox{trace}\sqrt{\sqrt{\rho_1}\rho_2 \sqrt{\rho_1}} 
\end{equation}
is doubly concave, i.e.,
\begin{equation}
G\Bigl(\lambda \rho_1 + (1 - \lambda) \sigma_1, 
\lambda \rho_2 + (1 - \lambda) \sigma_2\Bigr) \ge
\lambda G(\rho_1,\rho_2) + (1 - \lambda) G(\sigma_1,\sigma_2)\;.
\nonumber \\
\end{equation}

The proof uses a representation of the quantum fidelity in terms of  
measurement probabilities.  Given a measurement described by a 
positive-operator-valued measure (POVM) with POVM elements $E_i$, the 
probability for outcome $i$ is $p_i={\rm trace}\,\rho E_i$.  Fuchs and Caves 
\cite{Fuchs95a} showed that the quantum fidelity of $\rho_1$ and $\rho_2$ 
is the classical fidelity of the measurement probabilities for the 
measurement that, according to the classical fidelity, best distinguishes 
the two density operators, i.e., 
\begin{equation}
F(\rho_1,\rho_2)=\min_{\{E_i\}} F_{\rm cl}({\bf p}_1,{\bf p}_2)\;.
\label{Frep}
\end{equation}
Here the minimum is taken over all POVMs $\{E_i\}$, and ${\bf p}_1$ and
${\bf p}_2$ are the column vectors of measurement probabilities for 
$\rho_1$ and $\rho_2$ generated by the POVM $\{E_i\}$.

The proof begins by noting that for four positive real numbers,
\[ 0\le (\sqrt{x_1y_2}-\sqrt{x_2y_1})^2=x_1y_2+x_2y_1-2\sqrt{x_1x_2y_1y_2}\;, \]
from which it follows that the function $\sqrt{x_1x_2}$ is doubly concave, i.e.,
\begin{eqnarray}
\sqrt{[\lambda x_1+(1-\lambda)y_1][\lambda x_2+(1-\lambda)y_2]}
&=&\sqrt{\lambda^2 x_1x_2+(1-\lambda)^2y_1y_2+
\lambda(1-\lambda)(x_1y_2+x_2y_1)} \nonumber \\
&\ge&\sqrt{\lambda^2 x_1x_2+(1-\lambda)^2y_1y_2+
2\lambda(1-\lambda)\sqrt{x_1x_2y_1y_2}} \nonumber \\
&=&\lambda\sqrt{x_1x_2}+(1-\lambda)\sqrt{y_1y_2} \;. \nonumber
\end{eqnarray}

The square root of the classical fidelity,
\begin{equation}
G_{\rm cl}({\bf p},{\bf q})=\sqrt{F_{\rm cl}({\bf p},{\bf q})}
= \sum_{i=1}^{n} \sqrt{p_iq_i} \;,
\end{equation}
being a sum of such functions, is thus also doubly concave:
\begin{equation}
G_{\rm cl}
\Bigl(\lambda{\bf p}_1+(1-\lambda){\bf q}_1,
\lambda{\bf p}_2+(1-\lambda){\bf q}_2\Bigr)\ge
\lambda G_{\rm cl}({\bf p}_1,{\bf p}_2)
+(1-\lambda)G_{\rm cl}({\bf q}_1,{\bf q}_2)\;.
\end{equation}
Now use the representation~(\ref{Frep}), written in terms of square roots 
of fidelities, to show the double concavity of $G(\rho_1,\rho_2)$:
\begin{eqnarray}
G\Bigl(\lambda \rho_1 + (1 - \lambda) \sigma_1,
\lambda \rho_2 + (1 - \lambda) \sigma_2\Bigr)
&=& \min_{\{E_j\}} G_{\rm cl}
\Bigl(\lambda{\bf p}_1+(1-\lambda){\bf q}_1,
\lambda{\bf p}_2+(1-\lambda){\bf q}_2\Bigr) \nonumber \\
&\ge& \min_{\{E_j\}}
\,\Bigl(\lambda G_{\rm cl}({\bf p}_1,{\bf p}_2)
+(1-\lambda)G_{\rm cl}({\bf q}_1,{\bf q}_2)\Bigr) \nonumber \\
&\ge& \min_{\{E_j\}} \lambda G_{\rm cl}({\bf p}_1,{\bf p}_2)
+\min_{\{F_j\}}(1-\lambda)G_{\rm cl}({\bf q}_1,{\bf q}_2) \nonumber \\
&=& \lambda G(\rho_1,\rho_2) + (1 - \lambda) G(\sigma_1,\sigma_2)\;.
\end{eqnarray}

(Another proof, by M.~A. Nielsen \cite{Nielsen98z}, 
uses the relation of quantum
fidelity to purifications.)
\end{appendix}


\end{document}